\documentclass[apsrev,amsmath,amssymb,reprint,longbibliography]{revtex4-1}

\usepackage{graphicx}			
\usepackage{color}
\usepackage{placeins}

\usepackage{array}
\newcolumntype{L}[1]{>{\raggedright\let\newline\\\arraybackslash\hspace{0pt}}m{#1}}
\newcolumntype{C}[1]{>{\centering\let\newline\\\arraybackslash\hspace{0pt}}m{#1}}
\newcolumntype{R}[1]{>{\raggedleft\let\newline\\\arraybackslash\hspace{0pt}}m{#1}}

\begin{document}

\title[]{Atomistic structure learning}

\author{Mathias S. J{\o}rgensen}
\author{Henrik L. Mortensen}
\author{S{\o}ren A. Meldgaard}
\author{Esben L. Kolsbjerg}
\author{Thomas L. Jacobsen}
\author{Knud H. S{\o}rensen}
\author{Bj{\o}rk Hammer}
\email{hammer@phys.au.dk}

\affiliation{
Department of Physics and Astronomy, and Interdisciplinary Nanoscience Center (iNANO), Aarhus University, DK-8000 Aarhus C, Denmark.
}

\date{\today}

\begin{abstract}
\noindent \textbf{
One endeavour of modern physical chemistry is to use bottom-up 
approaches to design materials and drugs with desired properties.
Here we introduce an atomistic structure learning algorithm (ASLA) that utilizes a
convolutional neural network to build 2D compounds and layered structures atom by
atom. The algorithm takes no prior data or knowledge on atomic interactions 
but inquires a first-principles quantum mechanical program for
physical properties. Using reinforcement learning,
the algorithm accumulates knowledge of chemical compound space for a given number
and type of atoms and stores this in the neural network, ultimately learning the blueprint for
the optimal structural arrangement of the atoms for a given target property. ASLA
is demonstrated to work on diverse problems, including grain boundaries in graphene
sheets, organic compound formation and a surface oxide structure. This approach to
structure prediction is a first step toward direct manipulation of atoms with artificially
intelligent first principles computer codes.}
\end{abstract}

\maketitle

The discovery of new materials for energy capture and energy storage
and the design of new drugs with targeted pharmacological properties
stand out as the ultimate goals of contemporary chemistry
research and materials science \cite{goals1,goals2}. The design spaces for these problems
are immense \cite{huge_space} and do not lend themselves to brute force or
random sampling \cite{brute_force}. This has led to the introduction of screening strategies in
which computational chemistry is utilized to identify the more
promising candidates before experimental synthesis and
characterization is carried out \cite{screening}. Machine learning techniques have further been
introduced to speed up such screening approaches \cite{screening_by_prediction}. Unsupervised
machine learning may serve to categorize substances whereby further search
efforts can be focused on representative members of different categories \cite{screening_by_category}. Supervised machine learning may direct the searches even
further as properties of unknown materials are predicted based on the
properties of known materials \cite{prediction}. In recent years, there has been a move
toward even more powerful use of machine learning, namely the employment
of active learning schemes such as reinforcement learning \cite{reinforcement1, reinforcement2}.

In active learning, the computational search is self-directed. It
thereby becomes independent on both the availability of large datasets
as in big data approaches \cite{big_data} and on the formulation by the research team of
a valid guiding hypothesis. In the chemistry domain, the power of active learning has been
demonstrated in drug design. Schemes in which molecules may be built
from encoding strings, e.g.~the simplified molecular-input line-entry system (SMILES) strings, have led to artificial
intelligence driven protocols, where a neural network based on experience
spells the string for the most promising next drug to be tested \cite{smiles_based_drug_design1}. For instance, generative
adversarial networks (GANs) have been utilized for this \cite{gan_based_drug_design}.

The strength of reinforcement learning was
recently demonstrated by Google DeepMind in a domain accessible to the general public \cite{atari,go,go_zero}. Here Atari 2600 computer games and
classical board games such as Chess and Go were tackled with techniques from computer vision and image recognition
and performances were achieved that surpassed those of all previous
algorithms and human champions.

\begin{figure*}[t]
    \centering
    \includegraphics[width=1.0\textwidth]{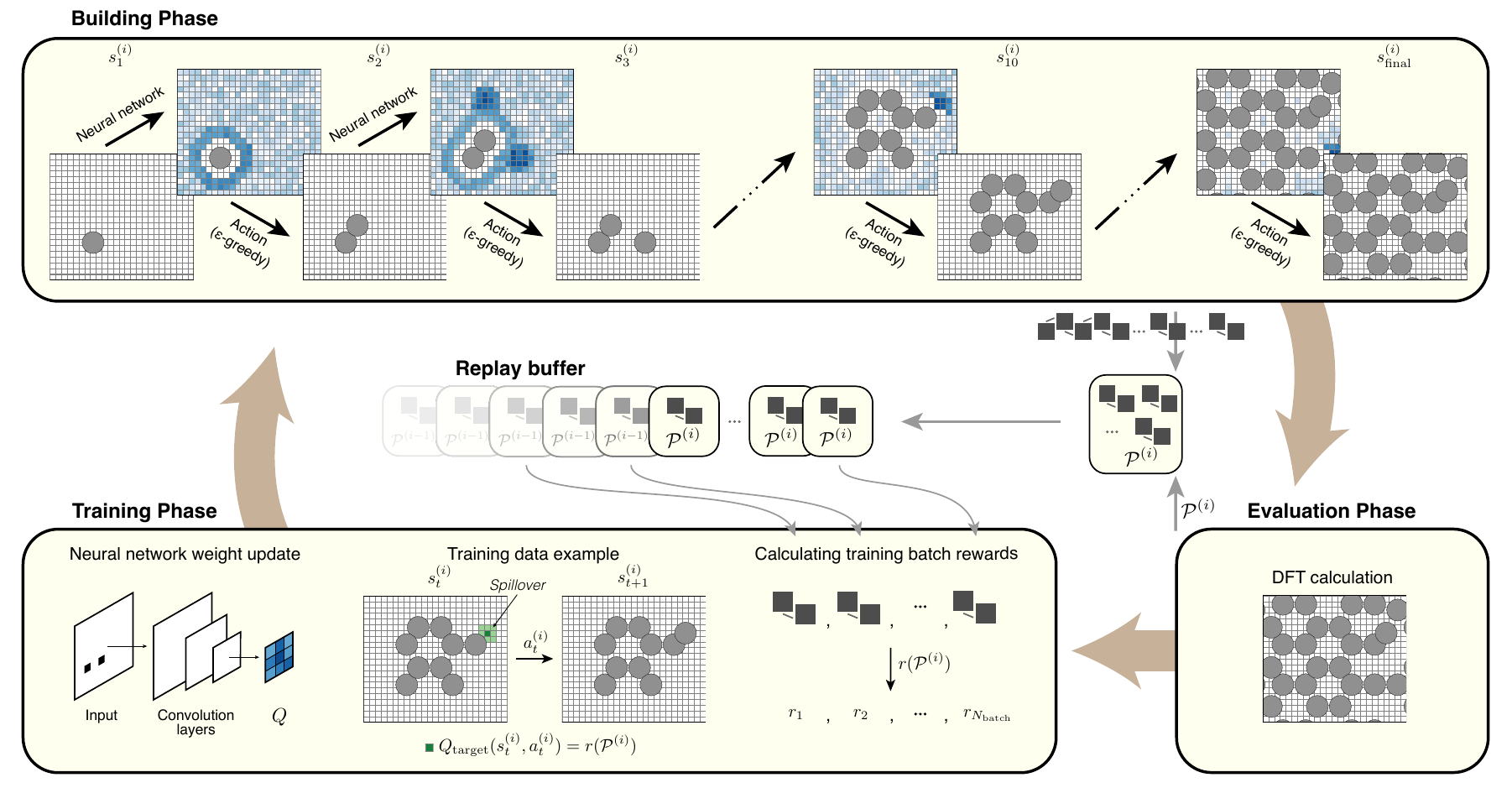}
    \caption{\textbf{The three phases of structure search in ASLA.} In each step $t$
      of the building phase for episode $i$, the atomistic structure $s^{(i)}_t$
      is input to the CNN that outputs $Q$-values (blue raster plot) estimating the expected reward for the final structure, $s^{(i)}_\mathrm{final}$. $N$ consecutive actions, $a^{(i)}_t$,
      are taken according to an $\varepsilon$-greedy policy thereby building the final structure.
      In the evaluation phase, a structure property, $\mathcal{P}^{(i)}$, is calculated.
      All episode steps are stored in the replay buffer from which a training batch is extracted. 
      Properties contained in the batch are then converted to rewards which represent $Q_\mathrm{target}$-values
      that the CNN is trained toward predicting.}
    \label{fig:flow}
\end{figure*}

\section{Atomistic reinforcement learning}

In the present work, we aim at bringing the image recognition approach into
the domain of quantum chemical guided materials search. I.e.~we
formulate a protocol for deep reinforcement learning in which a convolutional neural network (CNN) starts with zero knowledge and builds up its knowledge of
optimal organisation of matter via self-guided recurring inquiries for its produced
candidate structures. By optimal organisation of matter, we consider in this
first application primarily the thermodynamical stability of a
configuration as revealed by a quantum mechanical simulation.
The network is fed
a real-space image-like representation of the chemical structures and
is trained by "playing" against itself, striving to beat the
properties of its ever best candidate. As a result, the blueprint for favourable 
atomic arrangements is directly searched for and learned by the network. This deviates from existing approaches that based on intricate descriptors of the atomic environments \cite{huang, faber}
attempt to facilitate the calculation of
quantum mechanical energy landscapes at a low cost \cite{gap, behler, rupp, li, smith, chmiela} and use them in traditional global structure optimisation methods \cite{bh-nn, shang, esben}. Technical details on our CNN architecture
and training procedure are described in Methods.

The goal of an ASLA structure search is to generate structures of a
given stoichiometry which optimise a target structural property,
$\mathcal{P}$. This is done by repeatedly generating candidate
structures, evaluating a reward according to
$\mathcal{P}$, and learning from these experiences, or episodes, how
to generate the globally optimal structure. The three phases of an
ASLA structure search, \textit{building}, \textit{evaluation} and
\textit{training}, are depicted in Fig.~\ref{fig:flow}.

In the building phase, atoms are placed one at a time on a real space grid to form a structure candidate. $N$ atoms are placed in a sequence of \textit{actions} $a_t\in\{a_1, a_2, ..., a_N\}$. At each step of the candidate generation ASLA attempts to predict the expected reward of the final structure given the incomplete current structure. This information is stored in a CNN, which takes as input the incomplete, current configuration $s_t$, encoded in a simple one-hot tensor, and outputs the expected reward ($Q$-value) for each available action (see Extended Data Fig.~\ref{fig:CNN_architecture}). The action is chosen either as the one with the highest $Q$-value (greedy), or in $\varepsilon$-fraction of the time at random to encourage exploration. After the completion of a structure (episode $i$), the
property $\mathcal{P}^{(i)}$ of that final structure $s^{(i)}_\mathrm{final}$ is calculated in the evaluation phase, e.g.\ as the potential
energy according to density functional theory (DFT), $\mathcal{P}^{(i)}=E_\mathrm{DFT}(s^{(i)}_\mathrm{final})$, in which
case ASLA will eventually identify the thermodynamically most stable structure at $T=0$ K.

In the training phase, the CNN parameters are updated via
backpropagation, which lowers the root mean square
error between predicted $Q$-values and $Q_\mathrm{target}$-values
calculated according to:
\begin{equation}
Q_\mathrm{target}(s_t^{(i)},a_t^{(i)}) = r(\mathcal{P}^{(i)}),
\label{eq:bellman}
\end{equation}
where $r\in\left[-1,1\right]$ is the reward of completing episode $i$ with property
$\mathcal{P}^{(i)}$, where $1$ is
considered the better reward (reflecting e.g.\ the lowest potential
energy). Thus, the $Q$-value of action $a_t^{(i)}$ of
step $t$ is converged toward the reward following the properties of the \textit{final} configuration
$s_\mathrm{final}^{(i)}$, an approach also taken in
Monte Carlo reinforcement learning \cite{sutton}.  Equation \ref{eq:bellman} ensures that actions leading to
desired structures will be chosen more often. A training batch of $N_\mathrm{batch}$ state-action-property
tuples, $(s_t^{(i)},a_t^{(i)},\mathcal{P}^{(i)})$, is
selected from the most recent episode, the best episode, and via experience replay \cite{atari} of
older episode steps. The batch is augmented by rotation of structures on the grid. Rotational
equivariance is thus learned by the CNN instead of encoded in a molecular
descriptor as is done traditionally \cite{rupp, huang, schnet, faber}. This allows
the CNN to learn and operate directly on real space coordinates. We
furthermore take advantage of the smoothness of the potential energy
surface by smearing out the learning of $Q$-values to nearby grid
points (\textit{spillover}). See
Methods for details.

\begin{figure}[t]
    \centering
    \includegraphics[width=1.0\columnwidth]{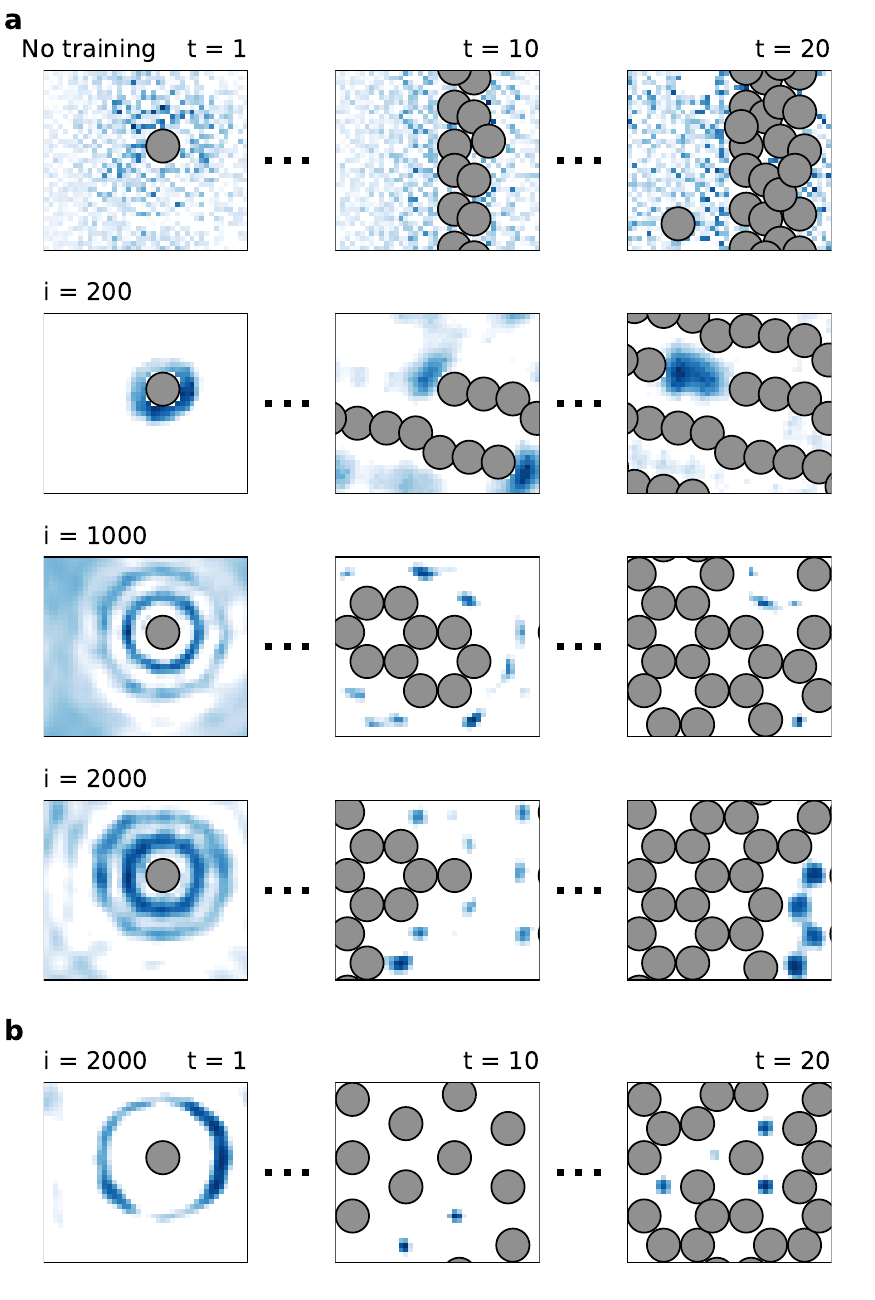}
    \caption{\textbf{ASLA learns to build a sheet of graphene.}
      \textbf{a}, One example of predicted $Q$-values for $t=1$, $10$ and $20$
      after 0, 200, 1000 and 2000 episodes respectively. \textbf{b}, An
      example of an agent which builds the globally optimal structure in
      a different sequence.}
    \label{fig:demo}
\end{figure}

\section{Building graphene}
As a first demonstration, an instance of ASLA (an \textit{agent}) learns how to
build a sheet of pristine graphene. One carbon atom is present in an
initial template structure, $s_{1}$, and $N=23$ carbon atoms are
placed by the agent on a grid with periodic boundary
conditions. Figure \ref{fig:demo}a presents one example of predicted
$Q$-values for $t=1, 10, 20$ during the building phase of one agent (for
statistics involving 50 independent agents, see Extended Data Fig.~\ref{fig:success}a+b).
Values are shown for the \textit{tabula rasa} agent (the \textit{untrained} agent)
and for the agent trained for 200, 1000 and 2000 episodes. At first, the
untrained CNN outputs random $Q$-values stemming from the random
initialization, and consequently a rather messy structure is
built. After 200 episodes, the agent has found the suboptimal strategy of
placing all the carbon atoms with reasonable interatomic distances,
but more or less on a straight line. The angular preference at this
point is thus to align all bonds.  After 1000 episodes, the agent has
realized that bond directions must alternate and now arrives at
building the honeycomb pattern of graphene.  After 2000
episodes, the agent still builds graphene, now in a slightly different
sequence.

The evolution over the episodes of the overall ring-shaped
patterns in the $Q$-values for $t=1$ show how the agent becomes aware
of the correct interatomic distances. The barely discernible six-fold
rotational symmetric modulation of the $Q$-values that develop within
the rings ends up being responsible for the agent choosing bond directions that
are compatible with accommodating graphene within the given periodic
boundary conditions.

Interestingly, the agent may discover the optimal structure through
different paths due to the random initialization and the stochastic
elements of the search. In Fig.~\ref{fig:demo}b, another agent has
evolved into building the graphene structure in a different
sequence. This time the agent has
learned to first place all next-nearest atoms (half of the structure) in a simple, regular pattern. The remaining atoms can then be placed in the same pattern but shifted. One can imagine how this pattern is easily encoded in the CNN. 

\section{Transfer learning}
\begin{figure}[ht]
  \includegraphics[width=\columnwidth]{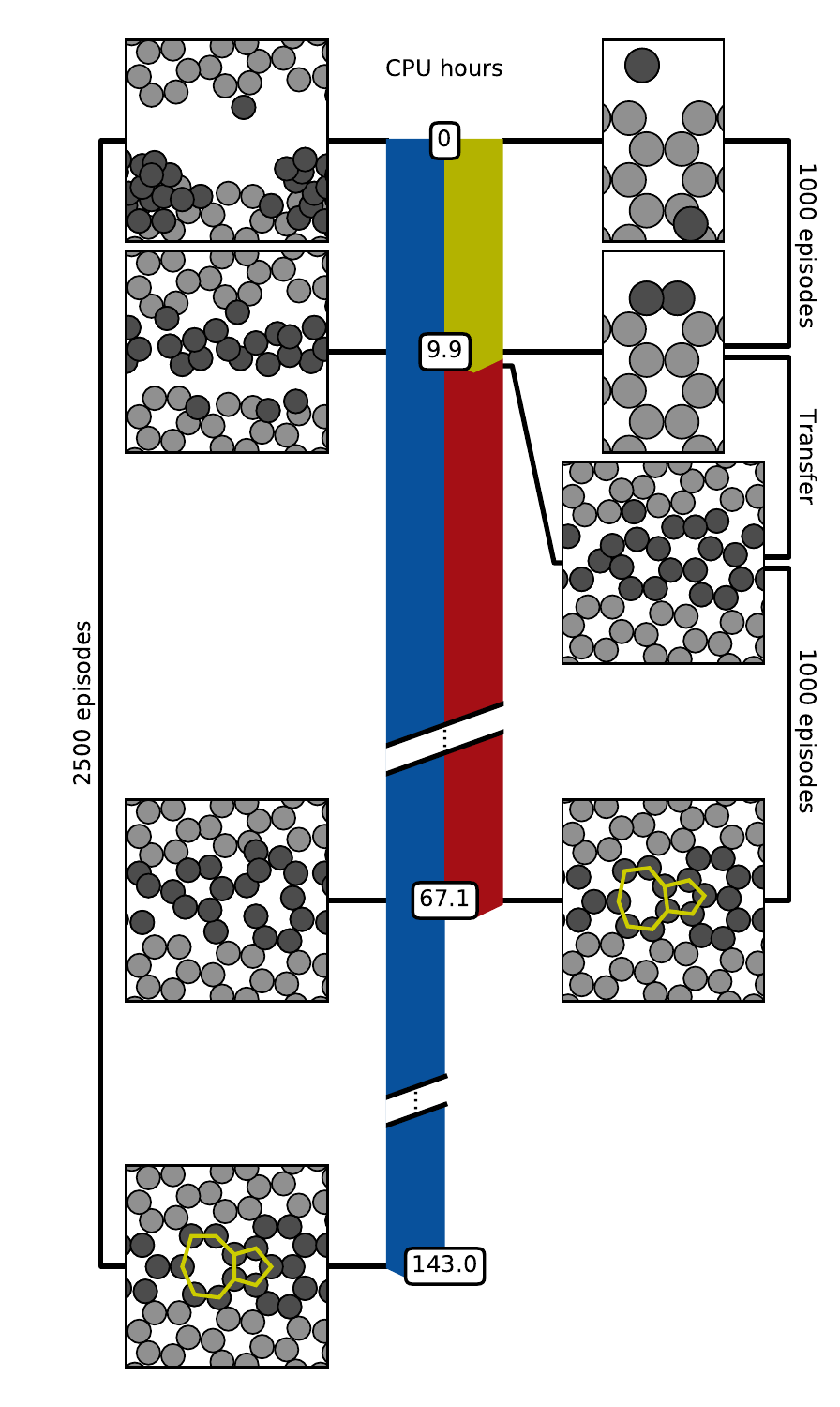}  
  \caption{\textbf{Demonstration of transfer learning}. CPU hours required to solve the graphene grain boundary problem without and with transfer learning (left and right part, respectively). The transferred agents pay a small amount of computational resources on the cheap graphene edge problem to get a head start on the grain boundary problem, where they solve it 50\% of the time using 57.2 CPU hours or less. However, the tabula rasa agents need 143.0 CPU hours to achieve the same.}
  \label{fig:transfer}
\end{figure}

\begin{figure}[ht]
  \includegraphics[width=\columnwidth]{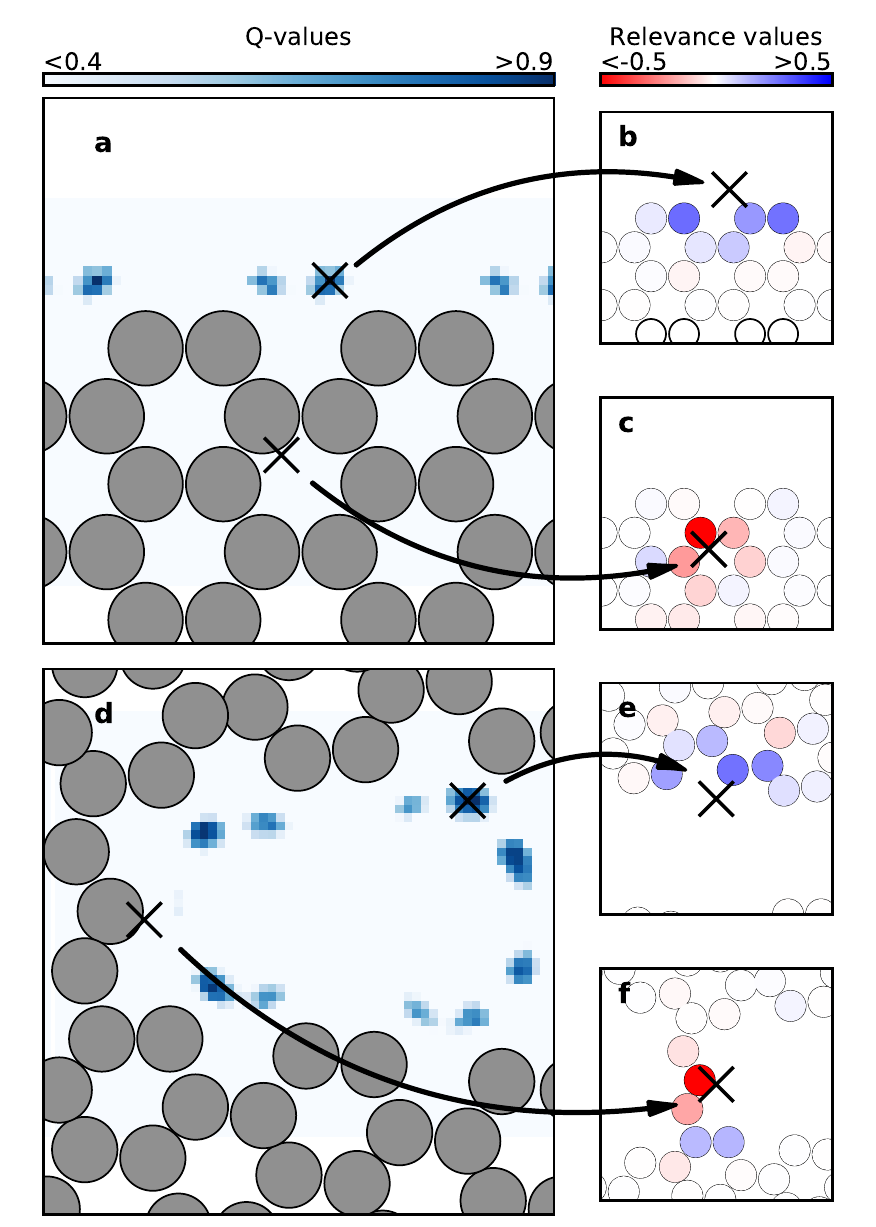}  
  \caption{\textbf{$Q$-values (left) and relevance values (right)
      given by an agent trained on the graphene edge}. \textbf{a}, The
    agent has learned that closing a six-membered carbon ring results
    in low potential energy and assigns high $Q$-values for these
    positions. \textbf{b}, Relevance values for the action ($\times$)
    reveal that nearby atoms are responsible for the high
    $Q$-values. \textbf{c}, If an atom is too close to an action
    ($\times$), this atom is responsible for low
    $Q$-values. \textbf{d}, The agent recognizes similar arrangements
    of atoms in the grain boundary (after 3 atoms have been added) and assigns high $Q$-values for
    corresponding actions. \textbf{e} and \textbf{f}, Although
    the agent has never seen this configuration, it applies the local
    knowledge learned in the graphene edge problem.}
  \label{fig:lrp}
\end{figure}

Moving on to a more complex problem, we consider a graphene grain
boundary which has been studied by Zhang et
al.~\cite{Zhang2015}. Here, an ASLA agent must place 20 atoms in
between two graphene sheets that are misaligned by 13.17$^\circ$, which induces a pentagon-heptagon (5-7) defect in the grain
boundary. The complete
structure under consideration is comprised of 80 atoms per cell. We first run an ensemble
of 50 tabula rasa agents. Within $\sim$2500 episodes, half of the 
agents have found the 5-7 defect structure as the optimal
structure (Fig.~\ref{fig:transfer} and Extended Data Fig.~\ref{fig:success}d).

\begin{figure}[t]
  \centering
  \includegraphics[width = \columnwidth]{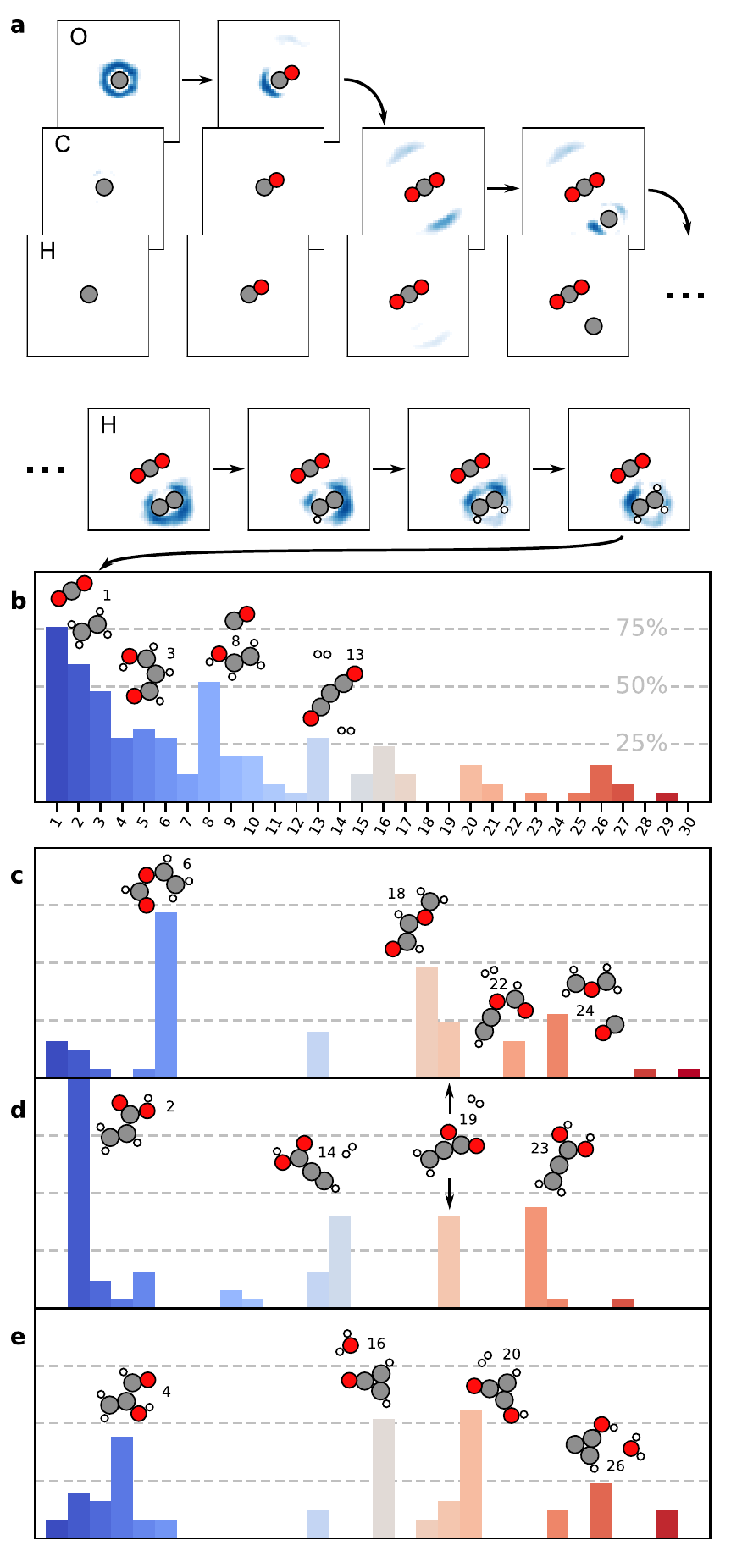}
  \caption{\textbf{Building C$_3$H$_4$O$_2$ compounds.} \textbf{a},
    Predicted $Q$-values in the process of building ethylene + carbon
    dioxide. \textbf{b}, Distribution of molecular compounds of
    stoichiometry C$_3$H$_4$O$_2$, sorted by formation energy, as found
    by 25 agents during 10000 episodes. For all structures see
    Extended Data Fig.~\ref{fig:30molecules}. The sum
    exceeds 100\% as a punishing term is employed once a build has been
    found repeatedly (see Methods).
\textbf{c}, Search results when rewarding O binding to 2$\times$C, coding for ether and epoxy groups.
\textbf{d}, Search results when rewarding C binding to C and 2$\times$O, coding for epoxy and acidic groups.
\textbf{e}, Search results when rewarding C binding to O and 2$\times$C, coding mainly for cyclic C$_3$ groups.}
  \label{fig:design}
\end{figure}

The number of required episodes can be lowered significantly with the use
of \textit{transfer learning}. Transfer learning is a widely used
approach in many domains where knowledge is acquired for one
problem and applied to another \cite{transfer_review_Pan_Yang, transfer_review_Taylor_Stone}.
Our CNN architecture allows us to utilize transfer learning, since each $Q$-value is evaluated based on the atomic-centered environment, which is independent of the cell size. The CNN may learn elements of how to discriminate favourable and unfavourable atomic arrangements in simple systems, and retain this knowledge when
used in larger, more computationally expensive systems.
To demonstrate this, we choose a simpler problem, namely that of placing two atoms at the edge of a graphene
nanoribbon (see upper right corner of Fig.~\ref{fig:transfer}).
Using our DFT setup this simple problem is cheaper to compute by a factor of $\sim$5 due to
the smaller number of atoms. The agents trained on this problem are transferred to build the grain
boundary. Although the graphene edge problem is too simple to
immediately infer the full solution to the grain boundary problem (Extended Data
Figs.~\ref{fig:sequence_transfer_no_training}+\ref{fig:sequence_transfer_with_training}), the pretrained agents hold a clear advantage over the tabula rasa agents. The average energies during the search reveal that the transferred agents produce structures close in energy
to the globally optimal structure even before further training (Extended Data Fig.~\ref{fig:success}f),
and now only about 1000 episodes are required for half of the agents to find the 5-7 defect.

To inspect the agents' motivation for predicting given $Q$-values, we
employ a variant of the layer-wise relevance propagation method (see
Methods). The relevance value of an atom indicates how influential
that atom was in the prediction of the $Q$-value, where 1 is positive influence, $-1$ is negative influence and $0$ is no influence. We show the $Q$- and relevance
values for the graphene edge and grain boundary problems in
Fig.~\ref{fig:lrp}, as given by one of the agents that has only been
trained on the graphene edge problem. In addition to solving the
graphene edge problem by closing a six-membered ring, the agent
recognizes similar local motifs (unclosed rings) in the grain boundary
problem which lead to favourable actions even without further training
(Fig.~\ref{fig:lrp}b+e). The relevance values in Fig.~\ref{fig:lrp}c+f
also reveal that the agent has learned not to place atoms too close no
matter the local environment.

\section{Molecular design}

The ability of ASLA to build multicomponent structures is easily
accommodated by adding to the CNN input a dimension representing
the atomic type. The CNN will then correspondingly output
a map of $Q$-values for each atomic type. When building a structure, a
greedy action now becomes that of choosing the next atom placement
according to the highest $Q$-value across all atomic types.

Figure \ref{fig:design}a illustrates the building strategy of a
self-trained ASLA agent assembling three carbon, four hydrogen and two
oxygen atoms into one or more molecules.  The agent identifies the
combination of a carbon dioxide and an ethylene molecule as its
preferred building motif. The reason for this build is the
thermodynamic preference for these molecules with the chosen
computational approach.  Depending on the initial conditions and the
stochastic elements of ASLA, we find that other agents may
train themselves into 
transiently building a plethora of other chemically meaningful
molecules given the present 2D constraint. The list of molecules
built includes acrylic acid, vinyl formate, vinyl alcohol,
formaldehyde, formic acid and many other well known chemicals (see
Extended Data Figs.~\ref{fig:molecule_sample}+\ref{fig:30molecules}).

Figure \ref{fig:design}b reveals the distribution of molecules
built when 25 independent agents are run for 10000 episodes, and
shows that ASLAs prolific production of different molecules has a
strong preference for the thermodynamically most stable compounds
(blue bars).  This is in accordance with the reward function depending
solely on the stability of the molecules.  We find, however, that the
building preference of ASLA may easily be nudged in particular
directions exploiting the bottom-up approach taken by the method. To
do this we modify the property function slightly (see Methods)
rewarding structures that contain some atomic motif. The histogram in
Fig.~\ref{fig:design}c shows the frequencies with which molecules are
built once the reward is increased for structures that have an oxygen
atom binding to \textit{two carbon atoms}. No constraints on the
precise bond lengths and angles are imposed, and ASLA consequently
produces both ethers (structures 6, 18, 22 and 
24) and an epoxy molecule (structure 19)
at high abundance, while the building of other
molecules is suppressed.

The production of the epoxy species intensifies as
evidenced by Fig.~\ref{fig:design}d once the rewarding motif becomes
that of a carbon atom bound to \textit{one carbon and two oxygen atoms}.
This motif further codes very strongly for acidic molecules
(structures 2, 14 and
23). In fact, now every agent
started eventually ends up building acrylic acid (structure 2) within the allowed number of episodes.
\FloatBarrier
\onecolumngrid
\begin{center}
  \centering
  \includegraphics[width = 1.0\textwidth]{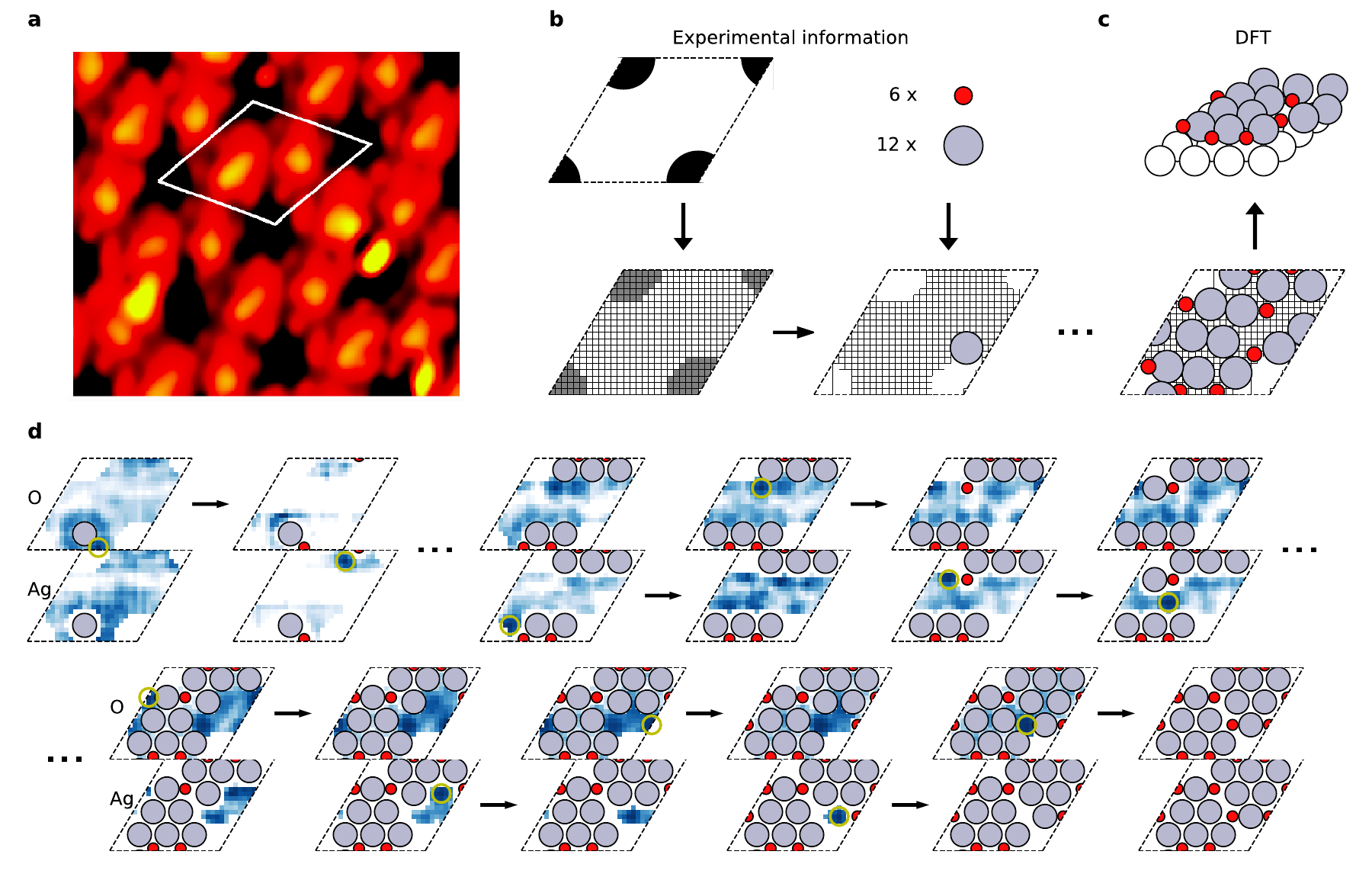}
\end{center}

\begin{figure*}[h!]
  \caption{\textbf{Learning to build a Ag$_{12}$O$_6$ overlayer
      structure}. \textbf{a}, Scanning tunneling microscopy
    image of the Ag(111)-\textit{p}(4 $\times$ 4)-O phase (adapted
    from \cite{ago_besenbacher}). \textbf{b}, Experimental information on
    super cell, absence of atoms in corner regions and stoichiometry
    is used as input to ASLA. \textbf{c}, Transfer of ASLAs 2D
    structure to become a surface layer in a DFT slab calculation.
    \textbf{d}, Predicted $Q$-values for an episode where the agent
    has self-learned to build the globally optimal overlayer structure.}
  \label{fig:ago} 
\end{figure*}
\FloatBarrier
\twocolumngrid

Switching to a
rewarding motif consisting of a carbon atom binding to
\textit{one oxygen and two carbon atoms} we find the results of
Fig.~\ref{fig:design}e.  Now, the search builds almost twice as many
2-hydroxyacrylaldehyde molecules (structure 4) and adopts a
clear preference for molecules with three carbon atoms in a ring, such as
the aromatic cyclopropenone and related molecules (structures
16, 20 and 26).

The ability of ASLA to direct its search in response to variations in the reward function represents a
simple illustration of how automated structure search may lead to design of molecules with certain properties.
The present approach may be used in search for larger organic compounds with some given backbone building blocks or some 
assemblies of functional groups. It might also be directly applicable when searching for organo-metalic compounds where the malleable
bonds mean that the desired motif may only be specified loosely as done here. However, for
searches that target e.g.\ ionization potentials, HOMO-LUMO gaps, deprotonation energies, and other global properties of molecules, we
envisage that more elaborate strategies for directing the searches must be implemented in order to truly conduct molecular design.

\section{Surface oxide on Ag(111)}

As a final, non-trivial example, 
the structure of oxidized silver, known to be an efficient catalyst for e.g.\ epoxidation
reactions \cite{ago_2005}, is considered. For this system, one particular surface oxide phase,
Ag(111)-$p(4\times 4)$-O, has been subject to many
studies since the 1970s resulting in several proposed structures
\cite{ago_2005, ago_1974, ago_2000, ago_besenbacher, ago_varga}. Key
experimental observations include STM images \cite{ago_2000,ago_besenbacher,ago_varga} (Fig.~\ref{fig:ago}a) of a regular rhombic surface unit cell containing 12 Ag atoms and 6 O
atoms per unit cell with void regions amounting to approximately 25\% of
the cell \cite{ago_varga}. Using this information, we prepare a grid
(Fig.~\ref{fig:ago}b) on which an agent can build a Ag$_{12}$O$_6$
overlayer structure. The property of a specific build is the DFT potential
energy of the overlayer structure once placed on a single, fixed Ag(111) layer representing the silver crystal
(Fig.~\ref{fig:ago}c). Employing transfer learning, starting from an agent trained in forming Ag$_4$O$_2$ clusters
(Extended Data Fig.~\ref{fig:sequence_AgO})
ASLA proves capable of learning by itself to
build the intricate structure
shown in Fig.~\ref{fig:ago}d.
In hindsight, 
this structure might appear highly intuitive to a trained inorganic surface chemist,
yet despite many efforts \cite{ago_2005} this correct structure remained elusive for
more than half a decade after the first characterization by STM imaging
in 2000 until it was finally identified in 2006 \cite{ago_besenbacher,ago_varga}.
It is
intriguing how such a puzzling structure may now emerge from a fully
autonomous reinforcement learning setup and a CNN of a
relatively modest size.

\FloatBarrier

\section{Outlook}
In the present work, we have demonstrated how deep reinforcement learning
can be used in conjunction with first principles computational methods
to automate the prediction of 2D molecular and inorganic compounds of
optimal stability. The work is readily
generalized into handling 3D atomistic structures by turning to 3D convolution
kernels and adding
one more spatial dimension to the input grid and to the output $Q$-value tensor.
This will increase the computation load of each $Q$-value prediction linearly in
the number of grid points added in the 3rd dimension, which is manageable. It
will, however, also open for more complex problems to be addressed which will
require more episodes to converge.

As more complex 2D or even 3D problems are tackled with the method, we foresee the need
for introducing further means of directing the global optimisation. The reinforcement
protocol will be slow in detecting structures that require several favourable
$\varepsilon$-moves to form and will suffer from conservatism, exploring mainly
close to previously built structures. Here, adding e.g.\ a beam search strategy, Monte Carlo tree search
and possibly combining the method with a property predictor network allowing for
the screening of several orders of magnitude more structures than those handled
with the first principles method may be viable strategies.

Finally, we note that agents trained to act on a variable number of
atoms are desirable as they will be more versatile and capable of
solving more diverse problems. Such agents may be obtained by
introducing chemical potentials of atomic species.
Notwithstanding such possible extensions, we consider the
present work to represent a first peek into a rich research direction
that will eventually bring about first principles
computer codes for intelligent, bottom-up design and direct manipulation of atoms 
in chemical physics and materials science.

\FloatBarrier

\bibliography{refs}

\noindent\textbf{Acknowledgements}
We acknowledge support from VILLUM
FONDEN (Investigator grant, project no.\ 16562).

\noindent\textbf{Author contributions}
All authors contributed to conceiving the project, formulating the algorithms
and writing the computer codes. KHS acted as scrum master.
MSJ, ELK and TLJ established the first proof-of-concept implementation.
MSJ conducted the pristine graphene calculations.
HLM did the transfer/layer-wise relevance propagation calculations.
SAM did the multi-component/design calculations, while BH did the AgO
calculations. MSJ, HLM, SAM and BH wrote the paper.

\section{Methods}

\noindent\textbf{Neural network architecture.}  In all presented applications, a
plain CNN with three hidden layers of each
ten filters implemented with Tensorflow \cite{tensorflow} is used (see Extended Data Fig.~\ref{fig:CNN_architecture}).
The number of weights in the network depends on the filter size, grid spacing and number of atomic species, but is independent on the computational cell size used and on the
number of actions taken. Hidden
layers are activated with leakyRELU functions, while the output is
activated with a tanh function to get $Q \in [-1,1]$. The input is a one-hot encoded grid (1 at atom positions, 0 elsewhere), and the correct
number of output $Q$-values (one per input grid point) is retained by
adding padding to the input, with appropriate size and boundary conditions. Other layer types, such
as a fully connected layer, would disrupt the spatial coherence
between input and output and are thus not used. The filter dimensions
depend on a user-chosen physical radius and the grid spacing. E.g.~a
filter of width 3 \AA\space on a grid with spacing 0.2 \AA\space
has $15 \times 15$ weights per convolution kernel.

\noindent\textbf{Search policy.}  An action is chosen with a modified
$\varepsilon$-greedy sampling of the $Q$-values (see Extended Data Fig.~\ref{fig:policy}). The highest $Q$-value is
chosen with $(1-\varepsilon)$ probability, and a pseudo-random $Q$-value
is chosen with $\varepsilon=\varepsilon_0+\varepsilon_\nu$ probability, where $\varepsilon\in\left[0,1\right]$. Two types of actions contribute to $\varepsilon$. There is an $\varepsilon_0$ probability that the action is chosen completely randomly, while the actions are chosen among a fraction $\nu$ of the
best $Q$-values with $\varepsilon_\nu$ probability (hence a pseudo-random action), where
$\nu\in\left[0,1\right]$. The selectivity
parameter $\nu$ is introduced to avoid excessive exploration of
actions estimated to result in low-reward structures. To ensure that greedy policy builds are included in the training, every $N_p=5$th episode is performed without $\varepsilon$-actions. Atoms cannot be placed within some minimum distance of other atoms (defined as a factor $c_1$ times the average of the involved covalent radii). This is to avoid numerical complications with the use of DFT implementations. A maximum distance, with factor $c_2$, is used for the C$_3$H$_4$O$_2$ system to avoid excessive placement of isolated atoms (Fig.\ \ref{fig:design}, $c_2=2.5$).

\noindent\textbf{Neural network training.}  In all results of the
present work, the training batch contains $N_\textrm{batch}$ episode steps before augmentation. Among these
are \textit{all} steps building the optimal structure identified
so far, and all steps building the most recent candidate structure. The
remaining episodes are picked at random from the entire replay buffer. If
any state-action pair appears multiple times in the replay buffer, only the
episode step with the property that leads to the highest reward may be extracted.

The reward function scales all batch property values linearly to the interval $[-1,1]$, where 1 is considered the better reward. If the property values span a large interval, all values below a fixed property value is set to $-1$ in order to maintain high resolution of rewards close to 1. This is done for the graphene systems and the C$_3$H$_4$O$_2$ compounds, where the lower limit of $-1$ is fixed at an energy which is $\Delta E$ higher (less stable) than that of the energy corresponding to a reward of 1.

Spillover is incorporated into the backpropagation cost function by
also minimizing the difference between $Q_\mathrm{target}$ and
predicted $Q$-values of grid points within some radius
$r_\mathrm{spillover}$ of the grid point corresponding to the action
$a_t$. In all of the present work, $r_\mathrm{spillover}=0.3$ \AA. The
complete cost function is thus
\begin{equation}
J=\frac{1}{m} \sum_{i=1}^m \left[ Q-Q_\mathrm{target} + \sum_j \eta_j (Q_j^\mathrm{spillover} - Q_\mathrm{target}) \right]^2,
\end{equation}
where $m$ is the augmented batch size, and $\eta_j$ is a weight factor determining the amount of spillover for a grid point $j$ within the radius $r_\mathrm{spillover}$. Thus, the predicted $Q$-values of adjacent grid points, $Q_j^\mathrm{spillover}$, are also converged toward the $Q_\mathrm{target}$ value, but at a different rate determined by the factor $\eta_j$ ($\eta_j=0.1$ in the present work). CNN parameters are optimised with the ADAM gradient descent optimiser.

Rotational equivariance is learned from augmenting the batch by
rotating produced configurations (see Extended Data Fig.~\ref{fig:symmetry}). Similarly, to enable some spillover for AgO where the grid spacing
used (0.5 {\AA}) was larger than $r_\mathrm{spillover}$
we augmented the batch by ($s^{(i)}_t$, $\tilde{a}^{(i)}_t$, $\mathcal{P}^{(i)}$), where
$\tilde{a}^{(i)}_t$ deviates from the actually tested action, $a^{(i)}_t$, by $\pm$ one pixel in the $x-$ and $y-$directions.

\noindent\textbf{Layer-wise relevance propagation.}  Layer-wise
relevance propagation seeks to decompose the CNN output, i.e.~the
$Q$-value, and distribute it onto the CNN input in a manner that
assigns a higher share, or \textit{relevance}, to more influential input
pixels (Fig.\ \ref{fig:lrp}). The outputs are distributed layer by layer, while
conserving relevance, such that the sum of the relevance values in
each layer equals the output. The relevances are computed by backwards
propagating the output through the CNN. We compute the relevance of
neuron $i$ in layer $l$ by

\begin{equation}
  \label{eq:lrp}
  R_i^{l} = \sum_j \frac{x_i w_{ij}}{\sum_i x_i w_{ij}+\tau} R_j^{l+1}.
\end{equation}
Here, $x_i$ is the activation of node $i$ connected to node $j$ by the
weight $w_{ij}$, while $\tau$ is a small number to avoid numerical
instability when the denominator becomes small. Equation \ref{eq:lrp}
distributes relevance of neuron $j$ to neuron $i$ in the preceding
layer according to the contribution to the pre-activation of neuron
$j$ from neuron $i$ weighted by the total pre-activation of neuron
$j$. Equation \ref{eq:lrp} implies that bias weights are not used whenever
relevance values need to be calculated (i.e. in the graphene grain 
boundary and edge systems). 

\noindent\textbf{Exploration via punishment.}
The agent might settle in a local optimum, where a highly improbable series of $\varepsilon$-moves are needed in order to continue exploring for other optima. To nudge the agent onwards we introduce a punishment which is activated when the agent builds the same structure repeatedly, indicating that it is stuck (similar to count-based exploration strategies in reinforcement learning \cite{penalty}). The punishment term is added to the property value:
\begin{equation}
\tilde{\mathcal{P}}^{(i)} = \mathcal{P}^{(i)}+ \sum_m A \exp\left(- \frac{|s^{(i)}_\mathrm{final}-s^{(m)}_\mathrm{final}|^2}{2\sigma^2}\right),
\end{equation}
where $m$ runs over structures, $s^{(m)}_\mathrm{final}$, to be punished (repeated structures), and where
$A$ and $\sigma$ are constants. The distance $|s^{(i)}_\mathrm{final}-s^{(m)}_\mathrm{final}|$ is evaluated using a
bag-of-bonds representation \cite{bob} of the structures. Both C$_3$H$_4$O$_2$ and the AgO system utilize this technique.

\noindent\textbf{Property function modification.}
To facilitate molecular design the property values are changed to
\begin{equation}
\mathcal{P}^{(i)}=E_\mathrm{DFT}(s^{(i)}_\mathrm{final})
\cdot f(s^{(i)}_\mathrm{final}),\\
\end{equation}
where $f$ artificially strengthens the potential energy for structures fulfilling some criterion $g$ according to:
$$
f(s) = 
\left\{\begin{array}{ll}
1+\lambda&\mathrm{if\ }s\mathrm{\ fullfills\ }g\\
1&\mathrm{otherwise.}
\end{array}
\right.
$$
For instance, $g$ may be the criterion that some oxygen atom in the structure
is connected to two carbon atoms no matter the angle (two atoms are considered connected if they are
within the sum of the covalent radii + 0.2 \AA). In the present work, $\lambda=0.2$ is used.

\noindent\textbf{DFT calculations.}  All calculations are carried out
with the grid-based projector augmented-wave (GPAW \cite{gpaw})
software package and the atomic simulation environment (ASE)
\cite{ase}. The Perdew-Burke-Ernzerhof (PBE) exchange-correlation
functional \cite{PBE} is used together with a linear combination of atomic
orbitals (LCAO) for the graphene, and AgO surface reconstruction,
and a finite difference method for the
C$_3$H$_4$O$_2$ compound. All systems are modelled in vacuum. For further details on the system cell and grid dimensions, see Extended Data Table \ref{tab:parameters}.

\noindent\textbf{Hardware.}  All calculations were done on a Linux cluster with
nodes having up to 36 CPU cores. Neural network and DFT calculations
were parallelized over the same set of cores on one node. The CPU times given for
the transfer learning problem presented in Fig.~\ref{fig:transfer} were recorded when 4 CPU cores were used.

\noindent\textbf{Hyperparameters.}
For all relevant hyperparameters, see Extended Data Table \ref{tab:parameters}.


\renewcommand{\figurename}{\textbf{Extended Data Figure}}
\renewcommand{\tablename}{\textbf{Extended Data Table}}
\setcounter{figure}{0}

\begin{figure*}[ht!]
  \centering
  \includegraphics[width = 1.0\textwidth]{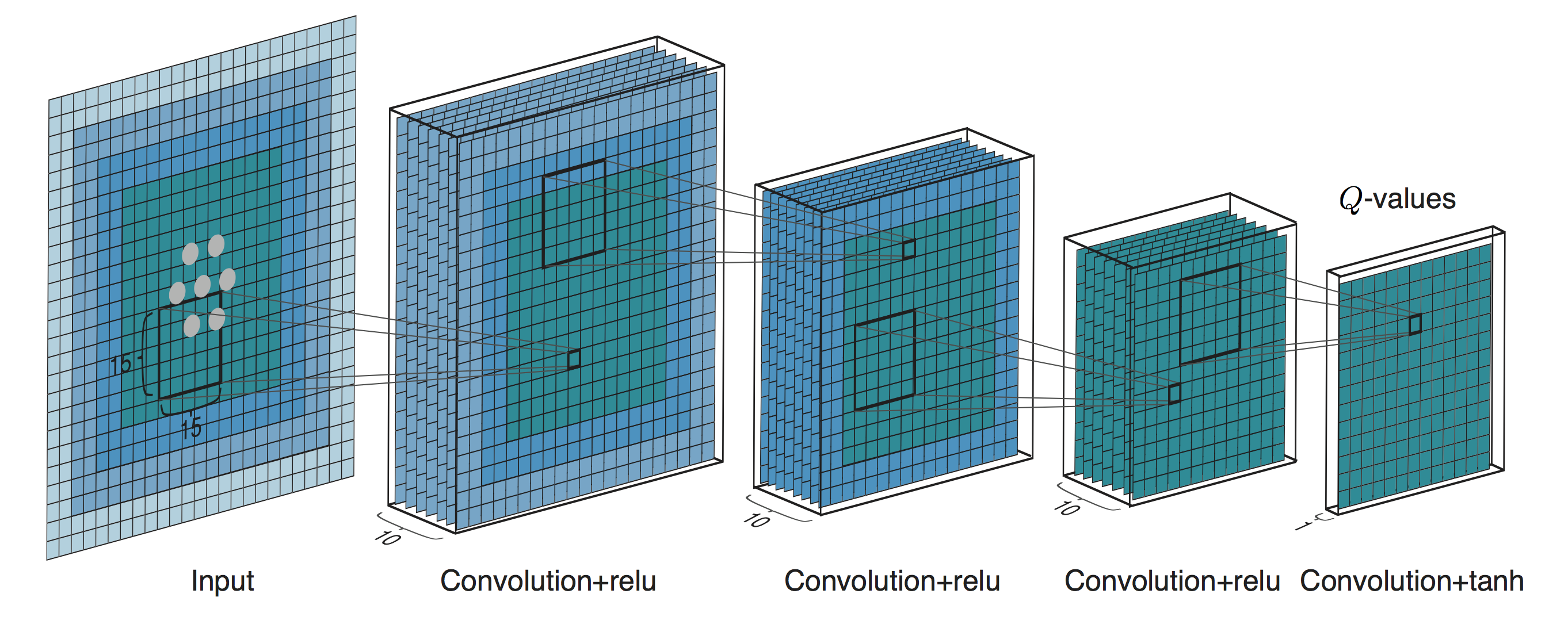}  
  \caption{\textbf{Layout of the CNN architecture for one atom species.} The input is atom positions mapped to a grid in a one-hot encoding with 1 at atom positions and 0 elsewhere. The input is forwarded through three convolution layers with 10 filters in each layer, arriving at the $Q$-values. We use convolution layers (i.e.~no pooling, fully-connected layers, etc.) such that we can map each $Q$-value to a real space coordinate. We pad the input appropriately, taking periodic boundary conditions into account if needed. Notice that the network accepts grids of arbitrary size since the output size scales accordingly. The input can readily be expanded in the third dimension to accommodate systems of several atomic species.}
  \label{fig:CNN_architecture}
\end{figure*}

\begin{figure*}[ht!]
  \centering
  \includegraphics[width = 1.0\textwidth]{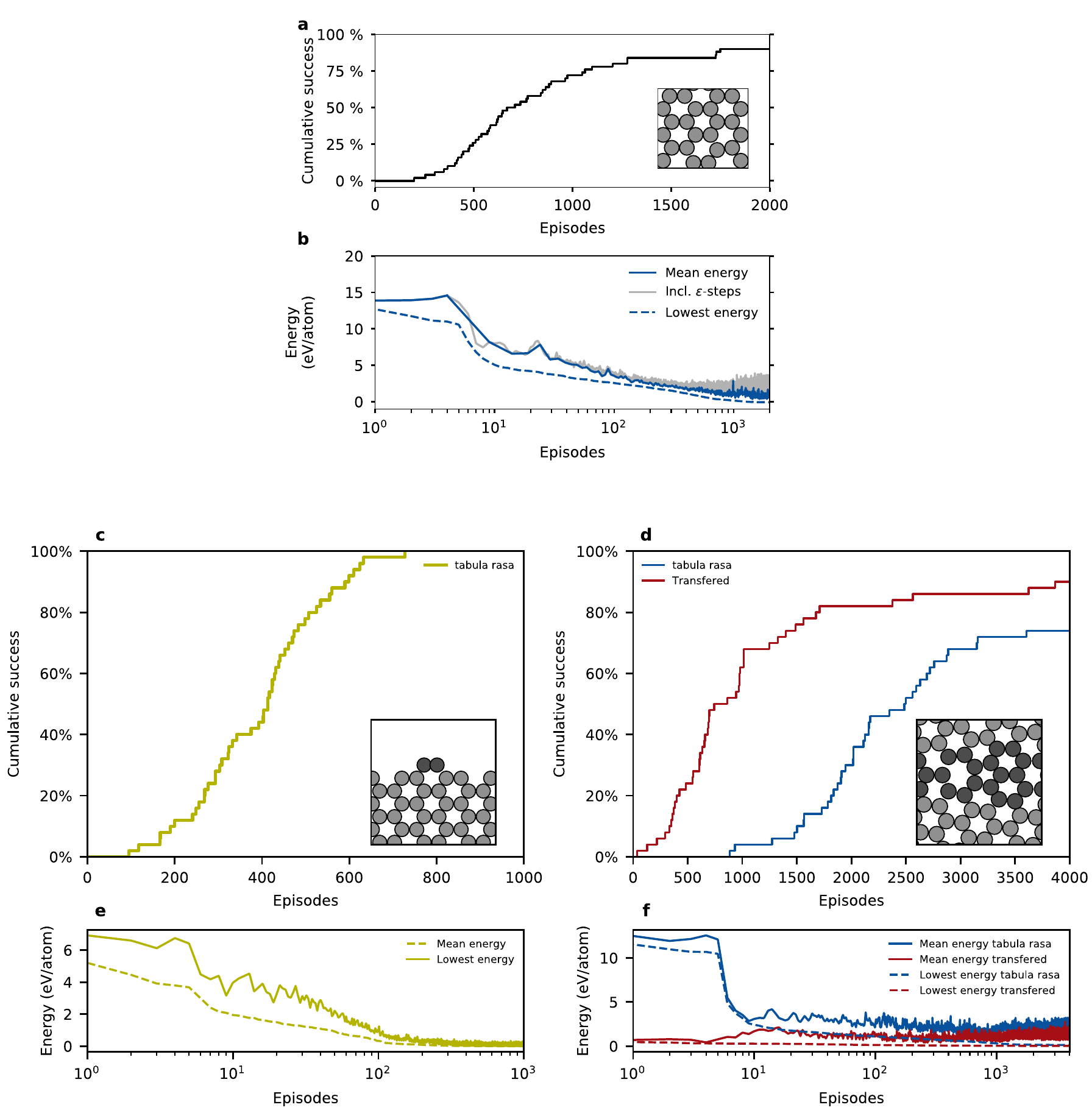}
  \caption{\textbf{ASLA searches for graphene systems.} \textbf{a}, The success of finding the structure of pristine graphene. After 2000 episodes with 50 independent ASLA agents, 90\% have found the globally optimal structure (inset) within 0.05 eV per atom. \textbf{b}, The energy of produced structures averaged over 50 agents relative to the energy of the globally optimal structure. The decrease in energy is most significant in the first few hundred episodes, where ASLA learns to place atoms at reasonable interatomic distances. \textbf{c}, The success of finding the structure of the graphene edge. After $\sim$700 episodes all agents (a total of 50) find the solution within 0.05 eV per atom. \textbf{d}, The success of finding the graphene grain boundary within 0.15 eV per atom, with and without transfer learning. An ensemble of 50 agents are started tabula rasa, while the 50 agents from the graphene edge problem are reused for transfer.}
  \label{fig:success}
\end{figure*}

\begin{figure*}[ht!]
  \centering
  \includegraphics[width = 0.9\textwidth]{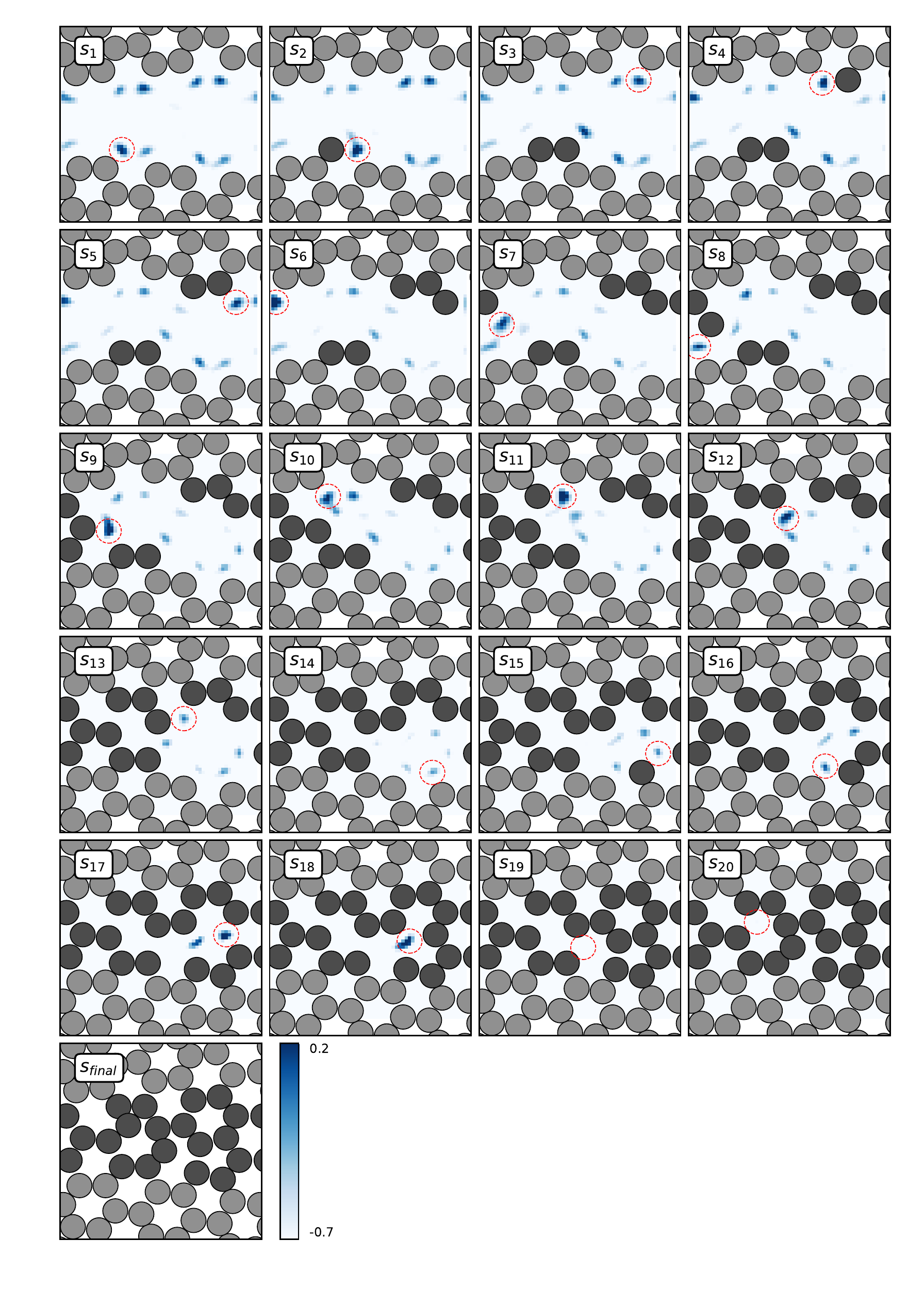}  
  \caption{\textbf{Transfered agent with no further training on the grain boundary problem.} This agent has only been trained on the graphene edge problem. $Q$-values are displayed in blue, and the red dotted circle indicates the position of the highest $Q$-value. The agent recognizes the similar environment between the two problems and builds a reasonable first guess. However, it does not know the complete solution.}
  \label{fig:sequence_transfer_no_training}
\end{figure*}

\begin{figure*}[ht!]
  \centering
  \includegraphics[width = 0.9 \textwidth]{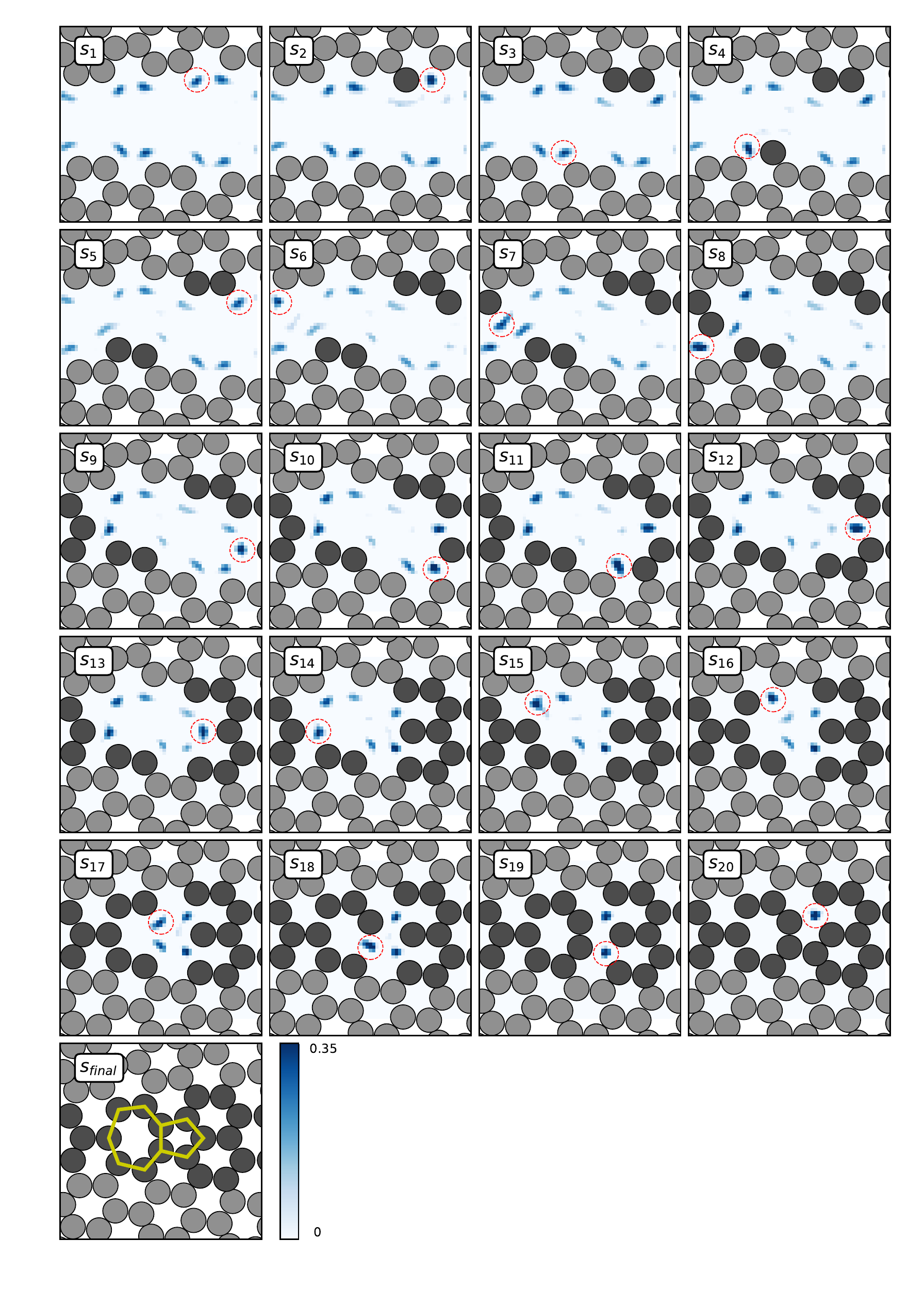}  
  \caption{\textbf{Transfered agent with further training on the grain boundary.} The same situation as in Extended Data Fig.~\ref{fig:sequence_transfer_no_training}, however, now the agent has been further trained on this problem for 555 episodes. The agent has learned to accommodate the 5-7 defect.}
  \label{fig:sequence_transfer_with_training}
\end{figure*}

\begin{figure*}[ht!]
  \centering
  \includegraphics[width = 1.0\textwidth]{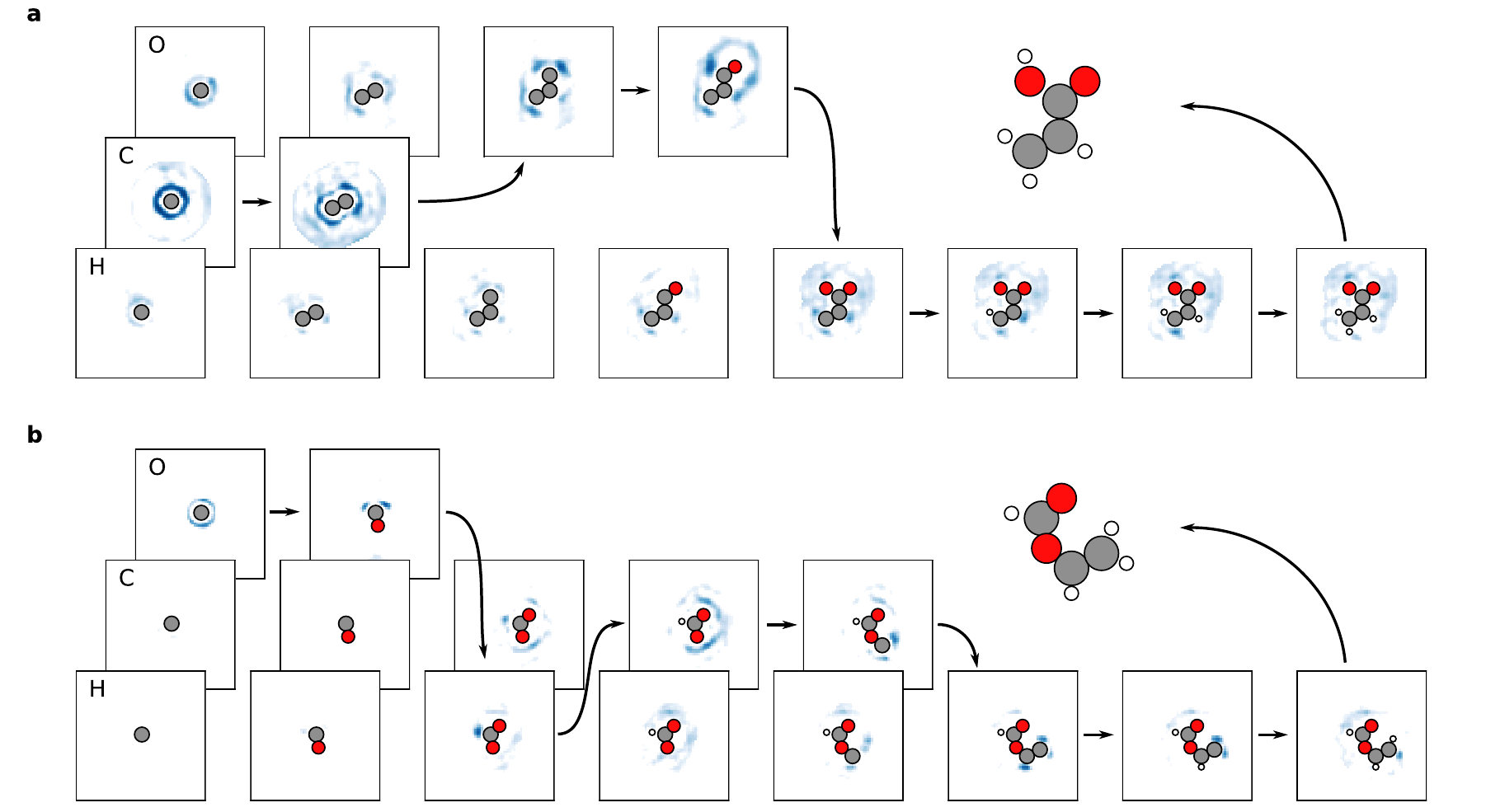}
  \caption{\textbf{Building a C$_3$H$_4$O$_2$ molecule.} \textbf{a}, Example of an acrylic acid build. In contrast to the carbon dioxide + ethylene structure, all carbon atoms are now placed prior to oxygen atoms. \textbf{b}, A vinyl formate build. A hydrogen atom is placed to build the formate before ASLA completes the structure by attaching the vinyl group.}
  \label{fig:molecule_sample}
\end{figure*}

\begin{figure*}[ht!]
  \centering
  \includegraphics[width = 1.0\textwidth]{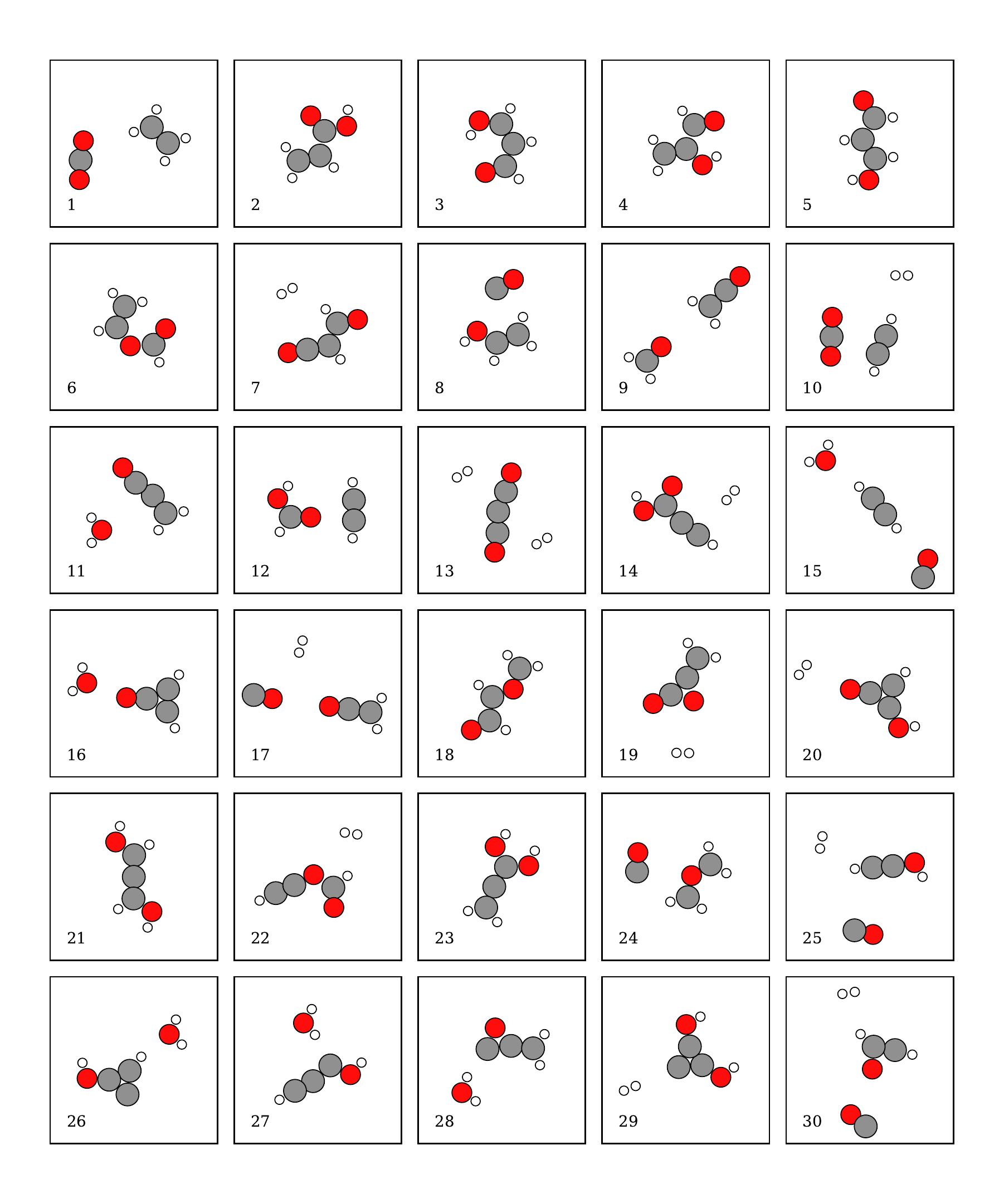}
  \caption{\textbf{Structures found for C$_3$H$_4$O$_2$.} The 30 most stable structures found 
    by the 4$\times$25 independent agents employed in
    Fig.\ \ref{fig:design}b-e. For isomers that may appear in several conformational forms,
    only one conformer is depicted. The structures shown have been relaxed without the use of the grid.
    Structures were counted in Fig.\ \ref{fig:design}b-e for greedy policy builds (every $N_p$'th episode) whenever (i) their graph was recognized as coinciding with those
    of the depicted structures, and (ii) their bond lengths and angles did not cause issues suggesting that they
    would not relax into the structures shown.}
  \label{fig:30molecules}
\end{figure*}

\begin{figure*}[ht!]
  \centering
  \includegraphics[width = 1.0\textwidth]{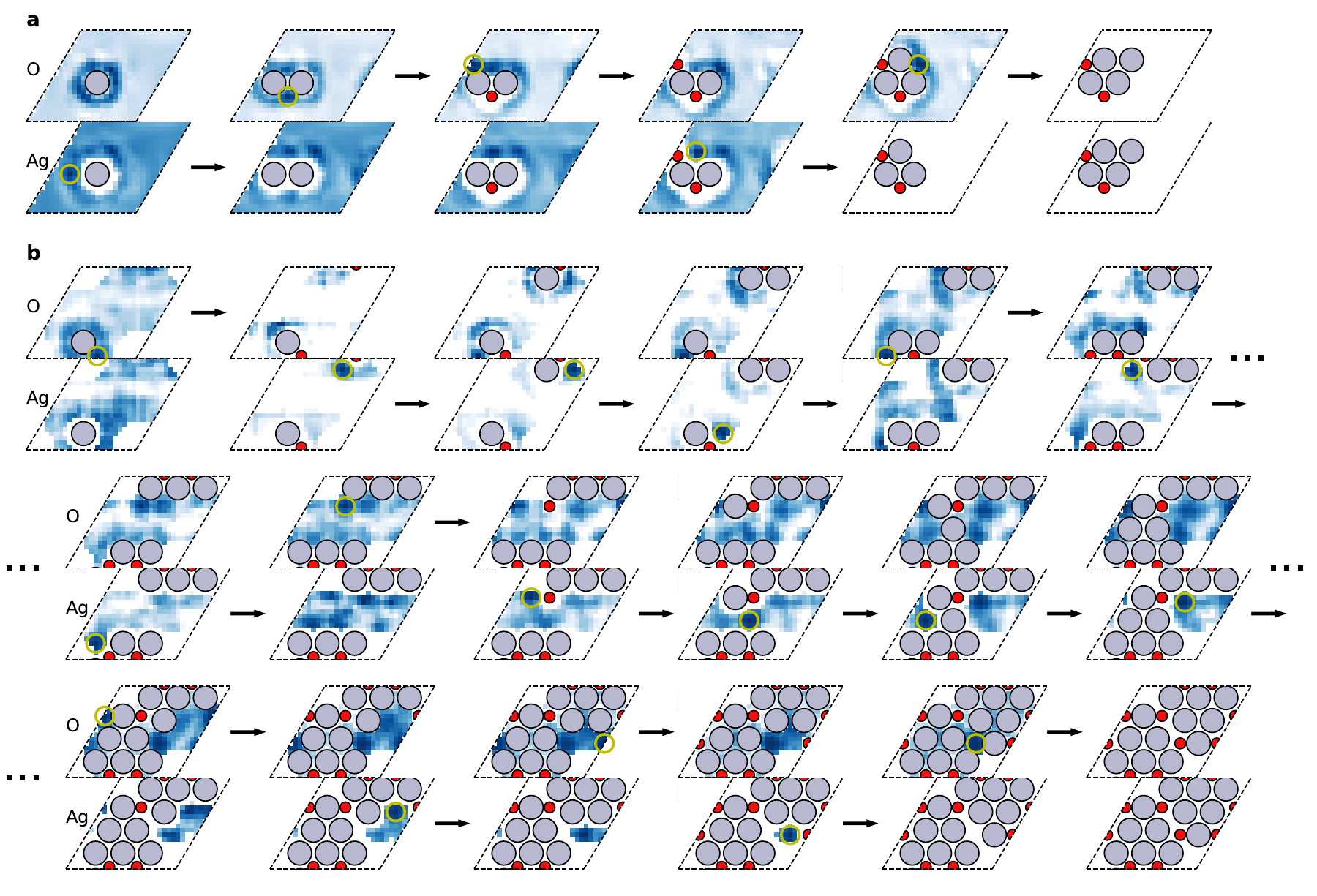}
  \caption{\textbf{Predicted $Q$-values for AgO structures}. Examples of full builds for \textbf{a}, A silver surface, Ag(111), supported Ag$_4$O$_2$ cluster structure and \textbf{b}, The Ag$_{12}$O$_6$ overlayer structure reported in \cite{ago_besenbacher,ago_varga}. The first Ag atom is placed at an \textsc{fcc} position with respect to the 2nd Ag layer. Other agents were trained for different choices of where the first Ag atom was placed but did not result in better structures. As in the entire work, the DFT total energy was evaluated without relaxation. The lateral separation of the two layers in the DFT calculation was chosen as 2.3 {\AA} and any buckling within the AgO overlayer was thus neglected.}
  \label{fig:sequence_AgO}
\end{figure*}

\begin{figure*}[ht!]
  \centering
  \includegraphics[width = 0.6\textwidth]{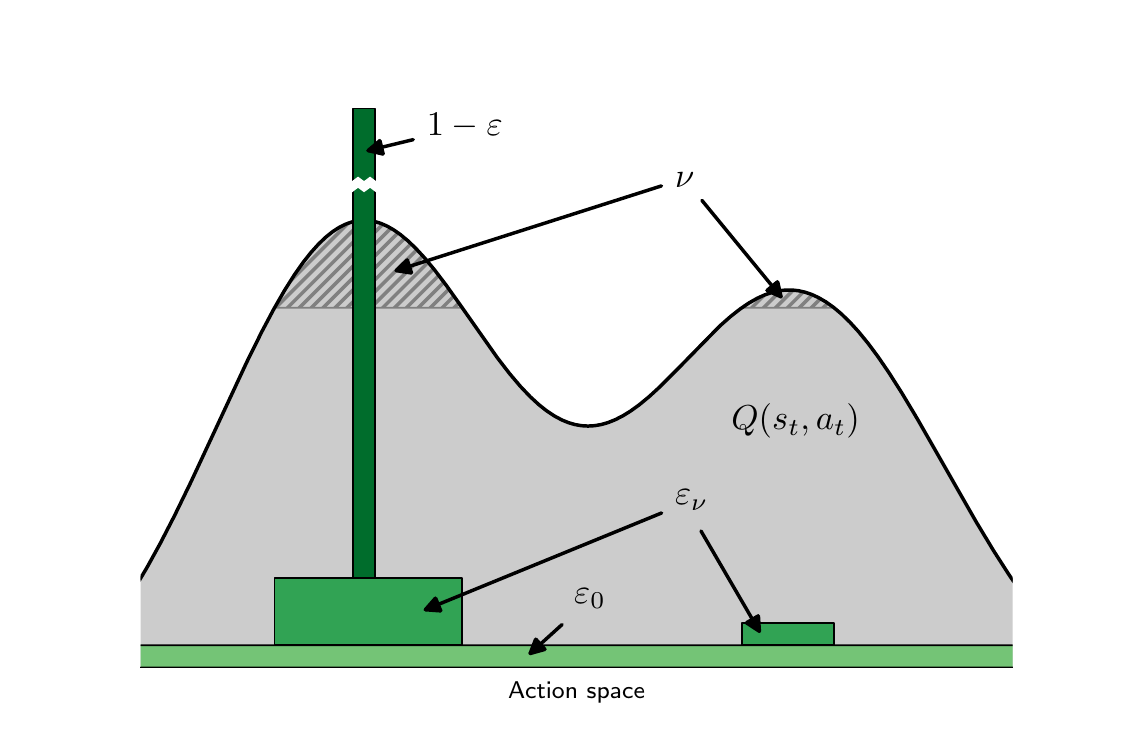}
  \caption{\textbf{Illustration of the pseudo-random $\varepsilon$-greedy policy.} From the entire space of actions, the action with the highest $Q$-value is chosen with $1-\varepsilon$ probability, while a pseudo-random action is chosen with $\varepsilon = \varepsilon_0 + \varepsilon_\nu$ probability. There is a probability $\varepsilon_0$ that it is chosen completely randomly, while the action is chosen among a fraction $\nu$ of the highest $Q$-values with $\varepsilon_\nu$ probability.} 
  \label{fig:policy}
\end{figure*}

\begin{figure*}[ht!]
  \centering
  \includegraphics[width = \textwidth]{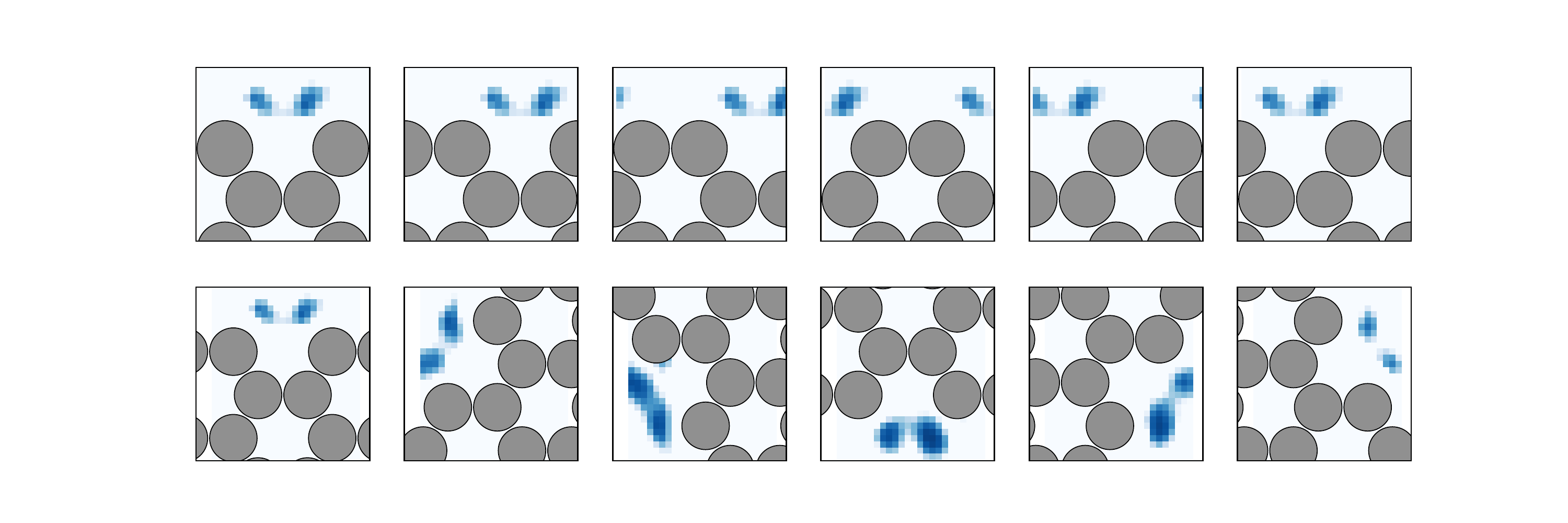}  
  \caption{\textbf{$Q$-values in the graphene edge problem for an agent trained for 700 episodes.} The top row demonstrates the translational equivariance of the $Q$-values directly inherited from the architecture of the CNN. The bottom row demonstrates the approximate rotational equivariance learned by the agent from augmenting the training data with rotated copies. }
  \label{fig:symmetry}
\end{figure*}

\clearpage

\begin{table*}[hb!]
\begin{center}
\parbox{14.4cm}{\caption{\textbf{Relevant hyperparameters and system dimensions of all systems in this work.}}\label{tab:parameters}}
\begin{tabular}{ L{3.6cm}  C{2.2cm}  C{2.2cm}  C{2.2cm}  C{2.2cm}  C{2.2cm} } \hline
  Hyperparameter & Graphene (pristine) & Graphene (edge) & Graphene (grain) & C$_3$H$_4$O$_2$ & AgO \\ \hline
  $\varepsilon$ 						& 0.10 & 0.70 & 0.10 & 0.20 & 0.125 \\
  $\varepsilon_\nu$ 				& 0.05 & 0.70 & 0.05 & 0.10 & 0.0625 \\
  $\nu$ 									& 0.02 & 0.05 & 0.02 & 0.10 & 0.02 \\
  $N_\textrm{batch}$ 			& 64 & 10 & 100 & 32 & 68 \\
  $\Delta E$ (eV/atom) 			& 2.0 & Not used & 0.5 & 3.3 & Not used \\
  $c_1$									& 0.5 & 0.4 & 0.4 & 0.6 & 0.65 \\
  $c_2$									& Not used & Not used & Not used & 2.5 & Not used \\
  Learning rate 						& $10^{-3}$ & $7.5\cdot10^{-4}$ & $10^{-3}$ & $10^{-3}$ & $10^{-3}$ \\
  Filter width (\AA) 				& 3 & 3 & 3 & 3 & 3 \\
  \# Conv.~layers 					& 3 & 3 & 3 & 3 & 3 \\
  \# Filters 							& 10 & 10 & 10 & 10 & 10 \\ \hline
  System dimensions 			&  &  &  &  &  \\ \hline
  Grid spacing (\AA)	   			& 0.20 & 0.18 & 0.18 & 0.20 & 0.50 \\
  Grid dimensions (points)		& $42\times37$ & $24\times40$ & $60\times50$ & $60\times60$ & $24\times20$ \\
  DFT cell dimensions (\AA) 	& $8.54\times7.40$ & $4.27\times7.12$ & $10.68\times8.90$ & $16.0\times16.0$ & $\left[\begin{array}{l}12\\0\end{array}\right]\times \left[\begin{array}{l}6.0\\10\end{array}\right]$  \\
  Periodic					 			& Yes (2D) & Yes (1D) & Yes (1D) & No & Yes (2D) \\ 
  $\mathbf{k}$-points					 			& $1\times1$ & 1 & 1 & N/A & $1\times1$ \\ \hline
\end{tabular}
\end{center}
\end{table*}

\end{document}